\documentclass[%
 reprint,
 amsmath,amssymb,
 aps,
]{revtex4-1}
\usepackage{graphicx}
\usepackage{dcolumn}
\usepackage{bm}

\begin{document}

\title{Power-efficiency-dissipation relations in linear thermodynamics}

\author{Karel Proesmans}
\email{Karel.Proesmans@Uhasselt.be}
\author{Bart Cleuren}%
\author{Christian Van den Broeck}
\affiliation{%
Hasselt University, B-3590 Diepenbeek, Belgium
}%

\date{\today}

\begin{abstract}
We derive general relations between maximum power, maximum efficiency, and minimum dissipation regimes from linear irreversible thermodynamics. The relations simplify further in the presence of a particular symmetry of the Onsager matrix, which can be derived from detailed balance. The results are illustrated on a periodically driven system and a three terminal device subject to an external magnetic field.
\end{abstract}

\maketitle


\section{Introduction}
Thermodynamic machines transform different forms of energy into one another. For such a machine, it would be of obvious interest to maximize the power $P$ and the efficiency $\eta$, and to minimize the dissipation $\dot{S}$ \cite{van2005thermodynamic,schmiedl2007efficiency,schmiedl2008efficiency,tu2008efficiency,rutten2009reaching,esposito2009universality,esposito2010efficiency,esposito_quantum-dot_2010,benenti2011thermodynamic,wang2012efficiency,izumida2012efficiency,golubeva2012efficiency,sheng2013universality,allahverdyan2013carnot,brandner2013strong,brandner2013multi,whitney2014most,PhysRevE.90.062140,izumida2014work,entin2014efficiency,holubec2014exactly,stark2014classical,jiang2014thermodynamic,polettini_efficiency_2015,cleuren2015universality,sheng2015constitutive,proesmans2015onsager,brandner2015thermodynamics,yamamoto2015thermodynamics,holubec2015efficiency,shiraishi2015attainability,raz2016geometric,ryabov2016maximum,shiraishi2016incompatibility,proesmans2015linear,bauer2016optimal,cerino2016Linear}.  The extrema (maximum or minimum) here are understood with respect to a variation of the engine's load parameters, which are often the ones that are easy to tune.  In general, the above goals are incompatible.  For example, the efficiency when operating at maximum power  is (in a time-symmetric setting) limited to half of  the reversible  efficiency $\eta_r=1$. The latter efficiency, being an overall upper bound, can only be reached  when operating reversibly, hence infinitely slowly. Consequently, the corresponding power vanishes. More generally, one may wonder whether there exist specific relationships between the regimes of maximum power (which will be denoted by a subscript $MP$),  maximum efficiency  (subscript $ME$) and minimum dissipation  (subscript $mD$). 
Recently, such relations have been discovered between $MP$ and $ME$ in the context of two case studies \cite{jiang2014thermodynamic,bauer2016optimal}. \\
In this letter, we derive general relations between the three regimes, within the framework of linear irreversible thermodynamics. Two results stand out. The first one is a remarkably simple relation linking $MP$ to $ME$:
\begin{equation}\eta_{MP}=\frac{P_{MP}}{2P_{MP}-P_{ME}}\eta_{ME}.\label{pedrel12}\end{equation}
As an implication note that, since power output $P_{ME}>0$ and efficiency $\eta_{MP}>0$ are positive, the efficiency at maximum power is at least half the maximum efficiency, $\eta_{MP}\geq\eta_{ME}/2$ .
The second result links the regimes of $MP$ and $mD$ by two equally simple equations:
 \begin{eqnarray}T\dot{S}_{mD}&=&\left(\frac{1}{\eta_{MP}}-\frac{1}{\eta_{ME}^2}-1\right){P_{MP}}+\frac{1}{\eta_{ME}^2}P_{ME},\label{pedrel11}\\P_{mD}&=&P_{MP}-\frac{1}{\eta_{ME}^2}\left(P_{MP}-P_{ME}\right),\label{pedrel13}\end{eqnarray}
where $T$ is the reference temperature of the system.
As a consequence, note that when minimum dissipation coincides with reversible operation, i.e., $\dot{S}_{mD}=0$ and $P_{mD}=0$,  one finds from Eqs.~(\ref{pedrel11},\ref{pedrel13}) that $\eta_{MP}=1/2$. \\
The above relations become more specific when the Onsager matrix, which links the thermodynamic fluxes and forces, satisfies a generalized Onsager symmetry condition, which we discuss  in more detail below.  The "standard" Onsager symmetry, which applies to time-symmetric machines, is a particular case.  Under this extra condition, the link between maximum power and efficiency, cf.~Eq.~(\ref{pedrel12}), splits into two separate relations, in agreement with the special cases discussed in \cite{jiang2014thermodynamic,bauer2016optimal}:
\begin{equation}\frac{P_{ME}}{P_{MP}}=1-\eta_{ME}^2,\;\;\;\eta_{MP}=\frac{\eta_{ME}}{1+\eta_{ME}^2},\label{pedrel22}\end{equation}
To mention some further implications of these results, reversible efficiency, $\eta_{ME}=1$ can only be reached when the power  goes to zero, $P_{ME}=0$. Furthermore, $0\leq\eta_{ME}\leq1$ implies $0\leq\eta_{MP}\leq1/2$, as first noted in \cite{van2005thermodynamic} (for a symmetric Onsager matrix).
 Note also that the equality sign in   $P_{ME}\leq P_{MP}$ is only reached for $\eta_{ME}=0$, hence $\eta_{MP}=0$, illustrating the conflict between maximizing efficiency and maximizing power.\\
Under the same generalized Onsager symmetry condition,  the links between maximum power and minimum dissipation, Eqs.~(\ref{pedrel11},\ref{pedrel13}), simplify as follows \footnote{Eq.~(49) from \cite{bauer2016optimal} is identical to the second part of our Eq.~(\ref{pedrel21}) provided one identifies
the 'idle heat flux' $J_{q}^{\textrm{idle}}$ with our minimal dissipation $\dot{S}_{mD}$.}:
\begin{equation}P_{mD}=0,\,\;\,T\dot{S}_{mD}=\left(\frac{1}{\eta_{MP}}-2\right)P_{MP}.\label{pedrel21}\end{equation}
Zero minimum dissipation (with $ P_{MP}  > 0$) implies $\eta_{MP}  = 1/ 2,$
$\eta_{ME}  = 1,$ and  $P_{ME}  = 0.$
Note the close interconnection between the results  Eqs.~(\ref{pedrel22},\ref{pedrel21}), since all of them follow from Eqs.~(\ref{pedrel12}-\ref{pedrel13}), if any one of them is valid.\\
We close the introduction with an important comment concerning the mathematical and physical content of the above relations. We will derive the above results first in the simple setting of two thermodynamic  fluxes and forces, linked by a two-by-two Onsager matrix $\bold{L}$. The relations Eqs.~(\ref{pedrel12}-\ref{pedrel13}) follow from straightforward algebra applied to the standard expressions from linear irreversible thermodynamics. No additional assumptions are needed. Eqs.~(\ref{pedrel22},\ref{pedrel21})  on the other hand require Onsager symmetry or anti-symmetry \footnote{ Onsager anti-symmetry is defined by $L_{ij}=-L_{ji}$ for all $i\neq j$} , i.e., ${L}_{12}=\pm {L}_{21}$. We next will show that both sets of results remain valid when the thermodynamic driving and loading force and flux are vectorial, i.e.,  they are composed of sub-forces and sub-fluxes, provided one performs the "full" optimization, i.e., with respect to all the components of the loading force. The validity of Eqs.~(\ref{pedrel22},\ref{pedrel21})  then rests in addition on a generalized Onsager symmetry, $ \bold{L}_{12}=\pm \bold{L}_{21}^{T}$  (T standing for the transpose) or $ \bold{L}_{12}=\pm \bold{L}_{21}$. This property can be derived from time-reversibility and detailed balance of the underlying micro-dynamics, and is therefore expected to have a very wide range of validity. We will illustrate this state of affairs on a system subject to a time-asymmetric periodic driving and a  three-terminal device with an external magnetic field. 
 
%

\section{Linear irreversible thermodynamics\label{GenFra}}
The thermodynamic processes that drive machines are generally induced by a spatial or temporal variation in quantities such as (inverse) temperature, chemical potential, pressure, etc. These differences are responsible for so-called thermodynamic forces, which we will denote by $F$. With every thermodynamic force, one can associate a flux, for example a heat flux or a particle flux, denoted as $J$.  The generic function of a machine is to transform one type of energy into another one. The simplest such construction thus features two forces, one playing the role of load force, say  $F_1$, and another functioning as driving force, $F_2$.
With proper definitions of fluxes and forces, the entropy production or dissipation, $\dot{S}$ can be written as a bilinear form \cite{prigogine1967introduction,de2013non}:
\begin{equation}\dot{S}=F_1J_1+F_2J_2.\label{Sdef}\end{equation}
The working regime is defined as a driving entropy producing  flux, say $J_2$ with $F_2J_2\geq0$, generating another flux $J_1$ against its own thermodynamic force $F_1J_1\leq 0$. The standard example is that of a thermal machine, where a downhill heat flux pushes particles up a potential. The quantities of interest are the net dissipation $\dot{S}$, given in (\ref{Sdef}), the power output $P$, which we define as \footnote{The temperature of the power producing device is the natural choice for the multiplicative factor  $T$. 
We however stress that the results  Eqs.~(\ref{pedrel12}-\ref{pedrel21}) are valid, irrespective of the choice of this temperature.}
\begin{equation}P=-TF_1J_1,\label{Pdef}\end{equation}
and the efficiency $\eta$,
\begin{equation}\eta=-\frac{F_1J_1}{F_2J_2}.\label{etadef}\end{equation}
The power output and efficiency are both positive by definition of the working regime. In addition, the second law $\dot{S}\geq0$ implies that, in the working regime, 
 $\eta\leq\eta_r=1$, with the reversible limit  $\eta=\eta_r$ reached for zero entropy production $\dot{S}=0$. Hence, one has:
 \begin{equation}
 \dot{S}\geq0,\;\;\;P\geq0,\;\;\;0\leq\eta\leq\eta_r=1.
 \end{equation}
Finally, by their definitions, power, efficiency and entropy production are not independent quantities but obey the following relation:
\begin{equation}\label{relation}
T\dot{S}=P\left(\frac{1}{\eta}-1\right).\end{equation}\\
Focusing on the regime of linear irreversible thermodynamics, one assumes that the thermodynamics forces are small, so that the associated thermodynamic fluxes are linear in the forces:
\begin{equation}\left(\begin{array}{c}
     J_1 \\
     J_2 
\end{array}\right)=\left(\begin{array}{cc}
     L_{11}& L_{12} \\
     L_{21} & L_{22}
\end{array}\right)\left(\begin{array}{c}
     F_1 \\
     F_2 
\end{array}\right).\end{equation}
The coefficients $L_{ij}$ are known as the Onsager coefficients.
For a given thermodynamic process, one can consider its time-inverse, denoted by a tilde. It is obtained by reversing the time dependencies and inverting the variables, such as speed and magnetic field, which are odd under time-inversion.  The above coefficients satisfy the so-called Onsager Casimir symmetry $\tilde{L}_{ij}=L_{ji},$
 \cite{casimir1945onsager}. This relation is particularly useful in the time-symmetric scenario with even variables, for which it reduces to the celebrated Onsager symmetry,
$L_{ij}=L_{ji}$ \cite{onsager1931reciprocal,onsager1931reciprocal2}.\\
We are now ready to calculate the values of the three key quantities, power, efficiency and dissipation, when performing the extremum of one of them with respect to the loading force $F_1$. In calculating the maximum efficiency and power, we will assume to be in the working regime.  This leads to nine expressions $P_{MP},P_{ME},P_{mD},\eta_{MP},\eta_{ME},\eta_{mD},\dot{S}_{MP},\dot{S}_{ME},\dot{S}_{mD}$, of which, in view of Eq.~(\ref{relation}), six are a-priori independent.  Straightforward algebra leads to the following explicit expressions:
\begin{equation}P_{MP}=T\frac{L_{12}^2F_2^2}{4L_{11}},\,\;\, \eta_{MP}=\frac{L_{12}^2}{4L_{11}L_{22}-2L_{12}L_{21}},\end{equation}
\begin{equation} P_{mD}=T\frac{(L_{12}^2-L_{21}^2)F_2^2}{4L_{11}},\,\;\, \dot{S}_{mD}=F_2^2\left(L_{22}-\frac{\left(L_{12}+L_{21}\right)^2}{4L_{11}}\right)\end{equation}
\begin{multline}P_{ME}=-TF_2^2\left(L_{11}L_{22}-\sqrt{L_{11}L_{22}\left(L_{11}L_{22}-L_{12}L_{21}\right)}\right)\\
\frac{\left(L_{11}L_{22}-L_{12}L_{21}-\sqrt{L_{11}L_{22}\left(L_{11}L_{22}-L_{12}L_{21}\right)}\right)}{L_{11}L_{21}^2}\end{multline}
\begin{multline}\eta_{ME}=-\left(L_{11}L_{22}-\sqrt{L_{11}L_{22}\left(L_{11}L_{22}-L_{12}L_{21}\right)}\right)\\
\frac{\left(L_{11}L_{22}-L_{12}L_{21}-\sqrt{L_{11}L_{22}\left(L_{11}L_{22}-L_{12}L_{21}\right)}\right)}{L_{21}^2\sqrt{L_{11}L_{22}\left(L_{11}L_{22}-L_{12}L_{21}\right)}}\end{multline}
The surprise  is that there are, in fact, only three independent quantities: one verifies by inspection the validity of the relations Eqs.~(\ref{pedrel12}-\ref{pedrel13}). 
In the case of Onsager symmetry or anti-symmetry, these equations further simplify with the appearance of one additional relation, cf.~Eqs.~(\ref{pedrel22},\ref{pedrel21}). Hence, we are left with only two independent quantities out of the original nine, for example any pair of power and efficiency,  $\dot{S}_{mD}$ and $\eta_{MP}$, $\dot{S}_{mD}$ and $P_{MP}$, etc..

\section{Multiple processes\label{MulSF}}
In a more general setting, a thermodynamic machine can involve many processes with input and output flux combinations of multiple sub-fluxes. 
Keeping the notation of subindices $i=1,2$ for loading and driving quantities respectively, the corresponding fluxes $\bold{J}_i$, forces $\bold{F}_i$ and Onsager coefficients $\bold{L}_{ij}$  are no longer scalars but vectors and matrices respectively. Onsager-Casimir symmetry predicts $\tilde{\bold{L}}_{ij}=\bold{L}_{ji}^T$.
Although the  proof now requires some more involved matrix algebra (cf.~supplemental materials), one can show that the first set of power-efficiency-dissipation relations, Eqs.~(\ref{pedrel12}-\ref{pedrel13}) remain valid provided the optimum is carried out with respect to all components of the loading force $\boldsymbol{F}_1$. Under the same optimization,  the second set of relations Eqs.~(\ref{pedrel22},\ref{pedrel21}) follows  for  Onsager matrices obeying the following generalized Onsager condition:
\begin{eqnarray}\bold{L}_{12;s}\bold{L}_{11;s}^{-1}\bold{L}_{12;s}&=&\bold{L}_{21;s}\bold{L}_{11;s}^{-1}\bold{L}_{21;s},\nonumber\\ \bold{L}_{12;a}\bold{L}_{11;s}^{-1}\bold{L}_{12;a}&=&\bold{L}_{21;a}\bold{L}_{11;s}^{-1}\bold{L}_{21;a}\label{gencon},\end{eqnarray}
with $\bold{L}_{ij;s}=(\bold{L}_{ij}+\bold{L}_{ij}^T)/2$, the symmetric part of the matrix and $\bold{L}_{ij;a}=(\bold{L}_{ij}-\bold{L}_{ij}^T)/2$ the anti-symmetric part of the matrix.
We make the important observation that this condition is satisfied for matrices obeying:
\begin{equation}\bold{L}_{12}=\pm \bold{L}_{21}^{T},\;\;\;\;\bold{L}_{12}=\pm \bold{L}_{21}\label{onsrel2bis}.\end{equation}
It is clear from Onsager symmetry that systems with time-symmetric driving satisfy this condition, but it may also hold for systems violating time-reversal symmetry.
Indeed, it has been shown that Onsager matrices of this form  arise as a consequence of detailed balance \cite{proesmans2015onsager,proesmans2015linear}, even though the set-up itself might break time-reversal symmetry, cf.~supplemental materials. Consequently, Eqs.~(\ref{pedrel22},\ref{pedrel21}) are expected to have a wide range of validity, including systems that break time-symmetry.
We stress again that the optimization needs to be carried out with respect to all components of the loading force. In the case of partial optimization, the corresponding effective Onsager matrix of lower rank  no longer satisfies Eq.~(\ref{gencon}), and therefore Eqs.~(\ref{pedrel22},\ref{pedrel21}) break down. On the other hand, Eqs.~(\ref{pedrel12}-\ref{pedrel13}) remain valid when the system is optimised with respect to the reduced set of variables, since the latter results are algebraic in nature, and do not require additional physical input.

\begin{figure}
    \includegraphics[width=0.7\columnwidth]{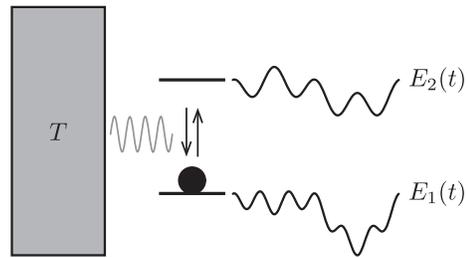}
    \caption{Schematic representation of a periodically driven two-level system in contact with a heat reservoir.}\label{fig:2lev}
\end{figure}
\section{Two examples\label{Twoex}}
We illustrate the above results on two systems that do not satisfy time-reversal symmetry:  a thermodynamic machine subject to  explicit time-periodic driving \cite{izumida2009onsager,izumida2010onsager,proesmans2015onsager,proesmans2015linear,brandner2015thermodynamics,bauer2016optimal,yamamoto2015thermodynamics,raz2015mimicking,falasco2015temperature,cerino2016Linear} and a three terminal device in an external magnetic field \cite{buttiker1988symmetry,sanchez2011thermoelectric,entin2012three,brandner2013strong,balachandran2013efficiency,Thieschmann,sanchez2015chiral,sanchez2015heat,hofer2015quantum}.

\begin{figure}
    \includegraphics[width=0.95\columnwidth]{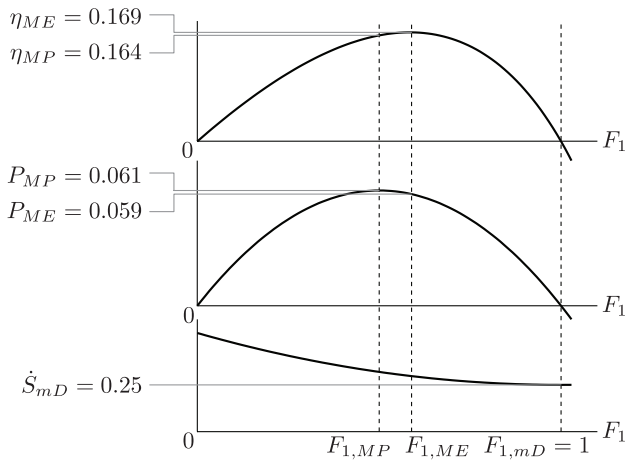}
\caption{Efficiency, power, and dissipation of a driven two-level system, $E_1(t)=F_1\cos(2\pi t/\mathcal{T})$ and $E_2(t)=F_2(\cos(2\pi t/\mathcal{T})+\cos(4\pi t/\mathcal{T}))$, with $\mathcal{T}=1$, $T=1$ and $F_2=1$. The Onsager coefficients are given by \cite{proesmans2015onsager}: $L_{11}=-L_{12}=-L_{21}=0.244$, $L_{22}=0.492$. $P_{mD}=0$, as can be seen by visual inspection. One also verifies
that $\eta_{ME}/(1+\eta_{ME}^2)=0.164=\eta_{MP}$, $P_{ME}/P_{MP}=0.971=1-\eta_{ME}^2$, $(1/\eta_{MP}-2)P_{MP}=0.25=\dot{S}_{mD}$, in agreement with Eqs.~(\ref{pedrel22},\ref{pedrel21}).}\label{fig:Lex}
\end{figure}

The first example is a  work-to-work converter consisting of a particle that can hop between two discrete energy levels, cf.~Fig.~\ref{fig:2lev}. Transitions are induced by a thermal bath, while the periodic modulation (period $\mathcal{T}$) of the energy levels via two external work mechanisms allows the conversion of work extracted from the second source, driving the second energy level, and delivered to the first source, loading the first energy level. The time dependence of the energy  in each level $i=1,2$ can be developed in terms of its Fourier components:
\begin{equation}E_i(t)=\sum_nF_{(i,n,s)}\sin\left(\frac{2\pi n t}{\mathcal{T}}\right)+F_{(i,n,c)}\cos\left(\frac{2\pi n t}{\mathcal{T}}\right),\end{equation}
where the amplitudes $F_{(i,n,c)}$ and $F_{(i,n,s)}$ play the role of thermodynamic forces, $n$ refers to the Fourier mode, and $c$ and $s$ to cosine and sine, respectively. Following standard techniques from stochastic thermodynamics \cite{harris_fluctuation_2007,sekimoto2010stochastic,seifert_stochastic_2012,SpinneyFord,van_den_broeck_ensemble_2014,tome2015stochastic}, one can determine the explicit expression for the elements of the associated Onsager matrix \cite{proesmans2015onsager} (cf.~supplemental materials):
\begin{equation}L_{(1,n,\sigma),(2,n,\sigma)}=-\frac{4\pi^3 n^3\left(4\pi^2n^2\bold{1}+\mathcal{T}^2{\bold{W}^{(0)}}^2\right)^{-1}_{12}p^{eq}_2}{\mathcal{T}},\end{equation}
where $\bold{W}^{(0)}$ and $\bold{p}^{eq}$ are the transition matrix and equilibrium probability distribution associated with the state of the particle in the absence of time-dependent driving, and $\sigma=s,c$. As a direct consequence of detailed balance, $\bold{W}^{(0)}_{12}\bold{p}^{eq}_2=\bold{W}^{(0)}_{21}\bold{p}^{eq}_1$,  one finds
\begin{equation}L_{(1,n,\sigma),(2,n,\sigma)}=L_{(2,n,\sigma),(1,n,\sigma)}.\end{equation}
Analogous relations are found for $L_{(1,n,\sigma),(2,m,\rho)}$, with $\rho\neq \sigma$ and $m\neq n$. We conclude that the following symmetry relation holds:
\begin{equation}L_{(1,n,\sigma),(2,m,\rho)}=L_{(2,n,\sigma),(1,m,\rho)},\end{equation}
which satisfies Eq.~(\ref{onsrel2bis}).
Hence, the second set of  power-efficiency-dissipation relations, Eq.~(\ref{pedrel22},\ref{pedrel21}) will be verified, see also \cite{bauer2016optimal} for a similar conclusion in a different model, and Fig.~\ref{fig:Lex} for an illustration in case of a
time-symmetric driving.

\begin{figure}
    \includegraphics[width=0.7\columnwidth]{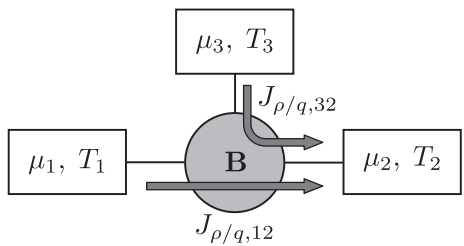}
    \caption{Schematic representation of the three terminal thermoelectric device.}\label{fig:3ter}
\end{figure}
As another example of a system with broken time-reversal symmetry we consider a three-terminal thermoelectric device in a magnetic field, cf.~Fig.~\ref{fig:3ter}. In this setup, three terminals are connected with each other via a central scattering region, inducing a particle flux $\bold{J}_\rho$ and a heat flux $\bold{J}_q$. In the working regime, the heat flux is from high to low temperature, while the particle flux is from low to high chemical potential. We assume that both fluxes are in the direction of the second reservoir in Fig.~\ref{fig:3ter}. In this way heat is converted into chemical energy. A magnetic field, $\bold{B}$ can be added to interact with the scattering region and break the time-reversal symmetry. An additional constraint which is often imposed is that the particle and heat flux through the third terminal vanish. The resulting $2\times 2$ Onsager matrix, associated with the heat and particle flux between reservoir $1$ and $2$, is generally not symmetric, and the efficiency at maximum power can reach values up to $\eta_{MP}=4/7$ \cite{brandner2013strong}, clearly violating the second set of power-efficiency-dissipation relations, Eqs.~(\ref{pedrel22},\ref{pedrel21}), cf.~supplemental materials. Crucial to this analysis, however, is the constraint that the fluxes through the third terminal are zero, which makes it impossible to fully optimize the power output.  Dropping the flux constraints will introduce thermodynamic sub-fluxes associated to the third terminal, and therefore $\bold{L}_{\rho q}$ and $\bold{L}_{q\rho}$ become $2\times 2$ matrices.\\
In the present context of linear thermodynamics, we set the reference values for temperature and chemical potential equal to those of the second reservoir, $T=T_2$ and $\mu=\mu_2$. The fluxes can be decomposed into a net flux from the first to the second terminal and from the third to the second terminal, $\bold{J}_{\rho/q}=(J_{\rho/q,12},J_{\rho/q,32})$ with associated thermodynamic forces $\bold{F}_\rho=e/T (\mu_1-\mu,\mu_3-\mu)$ and $\bold{F}_q=1/T^2 (T_1-T,T_3-T)$, where $e$ is the charge of one electron.
The behaviour of the central region is described by the scattering matrix, $\bold{S}(E,\bold{B})$, which gives the fluxes of electrons with energy $E$ between the different terminals, when an external magnetic field, $\bold{B}$, is applied to the central region.
The resulting Onsager matrix is given by \cite{sivan1986multichannel}:
\begin{equation}\bold{L}_{\alpha\beta}=\int^{\infty}_{-\infty}dE\,f_{\alpha\beta}(E)\left(\bold{1}-\bold{S}_{(1,3)}(E,\bold{B})\right)\label{LS},\end{equation}
with $\alpha,\beta=\rho$ or $q$, and $\bold{S}_{(1,3)}(E,\bold{B})$ the scattering matrix associated with the first and the third terminal only, and $f_{\alpha\beta}(E)$ a function independent of the central scattering region, and in particular of the presence of a magnetic field (cf.~supplemental materials). Hence, it is invariant under time-reversal symmetry and satisfies $f_{\rho q}(E)=f_{q\rho}(E)$, implying $\bold{L}_{\rho q}=\bold{L}_{q \rho}$. We conclude that Eqs.~(\ref{pedrel22},\ref{pedrel21}) will be valid when the optimization is carried out without constraints on the third terminal. In particular, the efficiency at maximum power will drop to a value below $1/2$.

\end{document}


\title{Power-efficiency-dissipation relations in linear thermodynamics}

\author{Karel Proesmans}
\email{Karel.Proesmans@Uhasselt.be}
\author{Bart Cleuren}%
\author{Christian Van den Broeck}
\affiliation{%
Hasselt University, theoretical physics
}%
\maketitle
\section{Derivation of power-efficiency-dissipation relation}
In this section, we show that the power-efficiency-dissipation relations mentioned in the main text remain valid when the load and driving fluxes and forces  become vectors, denoted by $\bold{J}_1$,  $\bold{J}_2$ and $\bold{F}_1$, $\bold{F}_2$, respectively. Note that the number of components of load and driving mechanism need not be the same. The Onsager matrix elements  linking them become (rectangular) matrices, notations $\bold{L}_{11},\bold{L}_{12},\bold{L}_{21},\bold{L}_{22}$:
\begin{equation}\left(\begin{array}{c}
     \bold{J}_1 \\
     \bold{J}_2 
\end{array}\right)=\left(\begin{array}{cc}
     \bold{L}_{11}& \bold{L}_{12} \\
     \bold{L}_{21} & \bold{L}_{22}
\end{array}\right)\left(\begin{array}{c}
     \bold{F}_1 \\
     \bold{F}_2 
\end{array}\right).\end{equation}

Power, efficiency and dissipation now read:
\begin{eqnarray}
P&=&-\bold{F}_1\bold{J}_1,\label{Pdefbold}\\
\eta&=&-\frac{\bold{F}_1\bold{J}_1}{\bold{F}_2\bold{J}_2}.\label{etadefbold}\\
\dot{S}&=&\bold{F}_1\bold{J}_1+\bold{F}_2\bold{J}_2\label{Sdefbold}.
 \end{eqnarray}
 As before, we want to evaluate  these quantities, and establish the relations between them, when performing the extemum  of one of them with respect to (all components of) the load force $\bold{F}_1$.  We start  by considering the regimes of maximum power and minimum dissipation. By setting the derivatives of power and dissipation with respect to $\bold{F}_1$ equal to zero, one finds:
\begin{equation}\bold{F}_{1;MP}=-\frac{1}{2}\bold{L}_{11;s}^{-1}\bold{L}_{12}\bold{F}_2,\end{equation}
\begin{equation}\bold{F}_{1;mD}=-\frac{1}{2}\bold{L}_{11;s}^{-1}\left(\bold{L}_{12}+\bold{L}^T_{21}\right)\bold{F}_2,\end{equation}
where we introduced the symmetric part of $\bold{L}_{11}$:
\begin{equation}\bold{L}_{11;s}=\frac{\bold{L}_{11}+\bold{L}_{11}^T}{2}.\end{equation} 
The products $\bold{F}_1\bold{J}_1$ and $\bold{F}_2\bold{J}_2$ in the MP and mD regimes thus become:
\begin{equation}\left(\bold{F}_1\bold{J}_1\right)_{MP}=-\frac{1}{4}\bold{F}^T_2\bold{L}_{12}^{T}\bold{L}_{11;s}^{-1}\bold{L}_{12}\bold{F}_2\label{F1def1}\end{equation}
\begin{equation}\left(\bold{F}_2\bold{J}_2\right)_{MP}=\bold{F}^T_2\left(\bold{L}_{22}-\frac{1}{2}\bold{L}_{21}\bold{L}_{11;s}^{-1}\bold{L}_{12}\right)\bold{F}_2\label{F1def2}\end{equation}
\begin{equation}\left(\bold{F}_1\bold{J}_1\right)_{mD}=\frac{1}{4}\bold{F}^T_2\left(\bold{L}_{12}^{T}+\bold{L}_{21}\right)\bold{L}_{11;s}^{-1}\left(\bold{L}_{21}^T-\bold{L}_{12}\right)\bold{F}_2\label{F1def3}\end{equation}
\begin{equation}\left(\bold{F}_2\bold{J}_2\right)_{mD}=\bold{F}^T_2\left(\bold{L}_{22}-\frac{1}{2}\bold{L}_{21}\bold{L}_{11;s}^{-1}\left(\bold{L}_{12}+\bold{L}_{21}^T\right)\right)\bold{F}_2\label{F1def4}\end{equation}

The expression for the load force, associated with to the regime of maximal efficiency, $\bold{F}_{1;ME}$, is more involved, see below Eq. (\ref{FME}). However, we can shortcircuit the calculation by noting that it satisfies the following equation:
\begin{equation}\left(\bold{F}_1\bold{J}_1\right)_{ME}\bold{L}_{21}^T\bold{F}_2=\left(\bold{F}_2\bold{J}_2\right)_{ME}\left(2\bold{L}_{11;s}\bold{F}_{1;ME}+\bold{L}_{12}\bold{F}_2\right).\label{F1def5}\end{equation}
To establish the link between maximum power and maximum efficiency, we multiply this last equation from the left with $\bold{F}_2^T\bold{L}_{12}^{T}\bold{L}_{11;s}^{-1}/2+\bold{F}^T_{1;ME}$. This gives:
\begin{multline}\left(\bold{F}_1\bold{J}_1\right)_{ME}\left(\bold{F}^T_{2}\bold{L}_{21;ME}\bold{F}_1+\frac{1}{2}\bold{F}_2\bold{L}_{21}\bold{L}_{11;s}^{-1}\bold{L}_{12}\bold{F}_2\right)\\=\left(\bold{F}_2\bold{J}_2\right)_{ME}\left(2\bold{F}^T_{1;ME}\bold{L}_{11;s}\bold{F}_{1;ME}+2\bold{F}^T_{1;ME}\bold{L}_{12}\bold{F}_2+\frac{1}{2}\bold{F}^T_2\bold{L}_{12}^{T}\bold{L}_{11;s}^{-1}\bold{L}_{12}\bold{F}_2\right)\label{supmaxef},\end{multline}
where we used the fact that a scalar variable is equal to its transpose, $\bold{F}^T_1\bold{A}^T\bold{F}_2=\bold{F}^T_2\bold{A}\bold{F}_1$. Using the above definitions of the fluxes, this can be rewritten as:
\begin{equation}\left(\bold{F}_1\bold{J}_1\right)_{ME}\left(\left(\bold{F}_2\bold{J}_2\right)_{ME}-\left(\bold{F}_2\bold{J}_2\right)_{MP}\right)=\left(\bold{F}_2\bold{J}_2\right)_{ME}\left(2\left(\bold{F}_1\bold{J}_1\right)_{ME}-2\left(\bold{F}_1\bold{J}_1\right)_{MP}\right).\end{equation}
The relation between maximum power and maximum efficiency, Eq.~(1) from the main text, follows by rewriting this equation in terms of power and efficiency Eqs. (\ref{Pdefbold},\ref{etadefbold}).

For the relations concerning minimal dissipation, we start with Eq.~(\ref{supmaxef}) and multiply from the left with $\bold{F}^T_{1;ME}+\bold{F}^T_2(\bold{L}_{12}^{T}+\bold{L}_{21})\bold{L}_{11;s}^{-1}/2$. After a short calculation, we arrive at:
\begin{multline}\left(\bold{F}_1\bold{J}_1\right)_{ME}\left(\left(\bold{F}_2\bold{J}_2\right)_{ME}-\left(\bold{F}_2\bold{J}_2\right)_{mD}\right)\\=\left(\bold{F}_2\bold{J}_2\right)_{ME}\left(2\left(\bold{F}_1\bold{J}_1\right)_{ME}-2\left(\bold{F}_1\bold{J}_1\right)_{MP}-\left(\bold{F}_2\bold{J}_2\right)_{MP}+\left(\bold{F}_2\bold{J}_2\right)_{ME}\right).\end{multline}
On the other hand, from Eqs.~(\ref{F1def1})-(\ref{F1def4}), we have
\begin{equation}2\left(\bold{F}_1\bold{J}_1\right)_{mD}+\left(\bold{F}_2\bold{J}_2\right)_{mD}=2\left(\bold{F}_1\bold{J}_1\right)_{MP}+\left(\bold{F}_2\bold{J}_2\right)_{MP}.\end{equation}
The results of the main text, Eqs.~(2,3), follow by rewriting the last two equations in terms of power, efficiency and dissipation.

For completeness, we give the explicit expression for $\bold{F}_{1;ME}$. It is found by multiplying Eq.~(\ref{F1def5}) with $\bold{F}^T_{2}$, and solving the resulting quadratic equation, leading to:
\begin{equation}\bold{F}_{1;ME}=-\frac{1}{2}\left(\bold{A}^{-1/2}\left(1-\left(1-4c\bold{A}^{1/2}\bold{B}^{-1}\left(\bold{B}^{T}\right)^{-1}\bold{A}^{1/2}\right)^{1/2}\right)\bold{A}^{-1/2}\bold{B}^T\right)\bold{F}_2,\label{FME}\end{equation}
\begin{eqnarray}\bold{A}&=&\bold{F}^T_2 \bold{L}_{21}\bold{F}_2\bold{L}_{11;s}-2\left(\bold{L}_{21}^{T}\bold{F}_2\bold{F}^T_2\bold{L}_{11;s}\right)_s,\\
\bold{B}&=&\bold{F}^T_2\bold{L}_{21}\bold{F}_2\bold{L}^{T}_{12}+2\bold{F}^T_2\bold{L}_{22;s}\bold{F}_2\bold{L}_{11;s}+\bold{F}^T_2\bold{L}_{12}\bold{F}_2\bold{L}_{21},\\
c&=&\bold{F}^T_2\bold{L}_{22;s}\bold{F}_2\bold{F}^T_2\bold{L}_{12}\bold{F}_2.\end{eqnarray}

To derive the second set of relations, we recall the important observation made in the main text that all equations hold if any one of them holds. Therefore, it is sufficient to investigate under which conditions one has that $P_{mD}=0$. This is the case, if and only if $\left(\bold{L}_{12}^{T}+\bold{L}_{21}\right)\bold{L}_{11;s}^{-1}\left(\bold{L}_{21}^T-\bold{L}_{12}\right)$ is fully anti-symmetric, which is equivalent with
\begin{equation}\bold{L}_{12;s}\bold{L}_{11;s}^{-1}\bold{L}_{12;s}=\bold{L}_{21;s}\bold{L}_{11;s}^{-1}\bold{L}_{21;s}\end{equation}
and
\begin{equation}\bold{L}_{12;a}\bold{L}_{11;s}^{-1}\bold{L}_{12;a}=\bold{L}_{21;a}\bold{L}_{11;s}^{-1}\bold{L}_{21;a},\end{equation}
where the subscripts $s$ and $a$ means that we are taking the symmetric or anti-symmetric part of the matrix.
A sufficient condition is that
\begin{equation}\bold{L}_{12;s}=\pm \bold{L}_{21;s},\end{equation}
and
\begin{equation}\bold{L}_{12;a}=\pm \bold{L}_{21;a},\end{equation}
which is equivalent with
\begin{equation}\bold{L}_{12}=\pm \bold{L}_{21}\end{equation}
or
\begin{equation}\bold{L}_{12}=\pm \bold{L}_{21}^T.\end{equation}

\section{Onsager coefficients for a periodically driven two-level system}
We shall now derive the Onsager coefficients of a periodically driven two-level system, in contact with a heat reservoir at temperature $T$. To do this we first write the time-dependencies of the energy levels as fourier series:
\begin{equation}E_i(t)=\sum_nF_{(i,n,s)}\sin\left(\frac{2\pi n t}{\mathcal{T}}\right)+F_{(i,n,c)}\cos\left(\frac{2\pi n t}{\mathcal{T}}\right),\end{equation}
for $i=1,2$. The average heat production per unit of time is given by
\begin{eqnarray}\dot{Q}&=&\frac{1}{\mathcal{T}}\int^{\mathcal{T}}_0dt\, E_1(t)\dot{p}_1(t)+\frac{1}{\mathcal{T}}\int^{\mathcal{T}}_0dt\, E_2(t)\dot{p}_2(t)\nonumber\\&=&\sum_n\frac{F_{(1,n,s)}}{\mathcal{T}}\int^{\mathcal{T}}_0dt\,\sin\left(\frac{2\pi n t}{\mathcal{T}}\right)\dot{p}_1(t)+\frac{F_{(1,n,c)}}{\mathcal{T}}\int^{\mathcal{T}}_0dt\,\cos\left(\frac{2\pi n t}{\mathcal{T}}\right)\dot{p}_1(t)\nonumber\\&&+\frac{F_{(2,n,s)}}{\mathcal{T}}\int^{\mathcal{T}}_0dt\,\sin\left(\frac{2\pi n t}{\mathcal{T}}\right)\dot{p}_1(t)+F_{(2,n,c)}\int^{\mathcal{T}}_0dt\,\cos\left(\frac{2\pi n t}{\mathcal{T}}\right)\dot{p}_2(t).\label{Qdef}\end{eqnarray}
Furthermore, we know that
\begin{equation}\dot{S}=\frac{\dot{Q}}{T},\end{equation}
and we assume that the entropy production can also be written in terms of thermodynamic forces and fluxes,
\begin{equation}\dot{S}=\sum_{\alpha}F_\alpha J_\alpha\label{Sdef}
\end{equation}
where the sum $\alpha$ runs over over all thermodynamic fluxes. Comparing Eq.~(\ref{Qdef}) with (\ref{Sdef}), we identify the natural choice for the force  $F_{(i,n,\sigma)}$, with $\sigma=s,c$, with corresponding thermodynamic fluxes given by
\begin{equation}J_{(i,n,s)}=\frac{1}{\mathcal{T}T}\int^{\mathcal{T}}_0dt\,\sin\left(\frac{2\pi n t}{\mathcal{T}}\right)\dot{p}_i(t),\label{Jdef}\end{equation}
and an analogous equation for $\sigma=c$.
To determine the Onsager coefficients we need to determine the probability distribution of the state of the particle at all times. This probability distribution can be found from the master equation
\begin{equation}
\dot{\bold{p}}(t)=\bold{W}(t)\bold{p}(t),\label{ME}
\end{equation}
where $\bold{W}(t)$ is the transition matrix. Furthermore, we define the adiabatic probability distribution $\bold{p}^{ad}(t)$ as the equilibrium distribution at time t:
\begin{equation}\bold{W}(t)\bold{p}^{ad}(t)=0\label{Ad}.\end{equation}
The elements of this distribution can be written as
\begin{equation}{p}^{ad}_i(t)=\frac{e^{-\frac{E_i(t)}{k_BT}}}{e^{-\frac{E_1(t)}{k_BT}}+e^{-\frac{E_2(t)}{k_BT}}}.\end{equation}
As we are only interested in the linear regime, we can expand the probability distribution and the transition matrix in terms of the thermodynamic forces:
\begin{equation}\bold{p}(t)=\bold{p}^{eq}+\sum_{i,n,\sigma}F_{(i,n,\sigma)}\bold{p}^{(1)}_{(i,n,\sigma)}(t)+O(F^2),\end{equation}
\begin{equation}\bold{p}^{ad}(t)=\bold{p}^{eq}+\sum_{i,n,\sigma}F_{(i,n,\sigma)}\bold{p}^{ad,(1)}_{(i,n,\sigma)}g_{(i,n,\sigma)}(t)+O(F^2),\end{equation}
\begin{equation}\bold{W}(t)=\bold{W}^{(0)}+\sum_{i,n,\sigma}F_{(i,n,\sigma)}\bold{W}^{(1)}_{(i,n,\sigma)}g_{(i,n,\sigma)}(t)+O(F^2),\end{equation}
where $\bold{p}^{eq}$ and $\bold{W}^{(0)}$ are the probability distribution and transition matrix in the absence of time-dependent driving, and
\begin{equation}
g_{(i,n,s)}(t)=\sin\left(\frac{2\pi nt}{\mathcal{T}}\right),\,\;\,g_{(i,n,c)}(t)=\cos\left(\frac{2\pi nt}{\mathcal{T}}\right).
\end{equation}
Expanding Eqs.~(\ref{ME}) and (\ref{Ad}) in terms of the thermodynamic forces gives:
\begin{equation}\dot{\bold{p}}^{(1)}_{(i,n,\sigma)}(t)=\bold{W}^{(0)}\bold{p}^{(1)}_{(i,n,\sigma)}(t)+\bold{W}^{(1)}_{(i,n,\sigma)}\bold{p}^{eq}g_{(i,n,\sigma)}(t),\end{equation}
\begin{equation}
\bold{W}^{(0)}\bold{p}^{ad,(1)}_{(i,n,\sigma)}+\bold{W}^{(1)}_{(i,n,\sigma)}\bold{p}^{eq}=0.
\end{equation}
Combining these equations leads to:
\begin{equation} \dot{\bold{p}}^{(1)}_{(i,n,\sigma)}(t)=\bold{W}^{(0)}\left(\bold{p}^{(1)}_{(i,n,\sigma)}(t)-\bold{p}^{ad,(1)}_{(i,n,\sigma)}g_{(i,n,\sigma)}(t)\right).
\end{equation}
This differential equation can be solved, leading to
\begin{equation}
\bold{p}(t)=\bold{p}^{ad}(t)-\int^{\infty}_0d\tau\, e^{\bold{W}^{(0)}\tau}\dot{\bold{p}}^{ad}(t-\tau)+O(F^2),
\end{equation}
which can be calculated exactly as the resulting probability distribution only depends on $\bold{p}^{ad}(t)$ and $\bold{W}^{(0)}$. Substitution in Eq.~(\ref{Jdef}), and solving the integrals gives the Onsager coefficients. These are given by:
\begin{equation} \bold{L}=\bigoplus_n \bold{L}_n,\end{equation}
with
\begin{equation}\bold{L}_n=\bold{L}_{n,1}+\bold{L}_{n,2},\end{equation}
\begin{eqnarray}\bold{L}_n^{(1)}&=& \bold{M}^{(1)}_n\otimes \left[ \begin{array}{cc}
1 & 0 \\
0 & 1 \end{array} \right]_{\sigma,\rho},\nonumber\\\bold{L}_n^{(2)}&=& \bold{M}^{(2)}_n\otimes \left[ \begin{array}{cc}
0 & 1 \\
-1 & 0 \end{array} \right]_{\sigma,\rho},\end{eqnarray}
\begin{eqnarray}\bold{M}^{(1)}_{n;(i,j)}&=&-2\pi^2 n^2\nonumber\\&& \left(\bold{W}^{(0)}\left(4\pi^2n^2\bold{1}+\mathcal{T}^2{\bold{W}^{(0)}}^2\right)^{-1}\right)_{i,j}p^{eq}_j,\nonumber\\
\bold{M}^{(2)}_{n;(i,j)}&=&\delta_{i,j}\frac{p^{eq}_k\pi n}{\mathcal{T}}\nonumber\\&&-\frac{4\pi^3 n^3\left(4\pi^2n^2\bold{1}+\mathcal{T}^2{\bold{W}^{(0)}}^2\right)^{-1}_{i,j}p^{eq}_j}{\mathcal{T}},\label{MFORMULA}
\end{eqnarray}
with $\sigma,\rho=s,c$ and $p_{k}^{eq}$ the equilibrium probability distribution associated with the two-level system in the absence of external driving and ${W}^{(0)}$ the transition matrix in the absence of driving. From this, we can derive
\begin{equation} \bold{L}_{ij}=\bigoplus_n \bold{L}_{n;ij}\end{equation}
with
\begin{equation}\bold{L}_{n;ij}=\left[\begin{array}{cc}
    {M}^{(1)}_{n;(i,j)} & {M}^{(2)}_{n;(i,j)} \\
    -{M}^{(2)}_{n;(i,j)} & {M}^{(1)}_{n;(i,j)}
\end{array}\right].\end{equation}
Due to the fact that this system satisfies detailed balance, we have
\begin{equation}W^{(0)}_{12}p^{eq}_2=W^{(0)}_{21}p^{eq}_1.\end{equation}
From the definitions of the $M$ matrices we than see
\begin{equation} {M}^{(1)}_{n;(1,2)}={M}^{(1)}_{n;(2,1)},\,\;\, {M}^{(2)}_{n;(1,2)}={M}^{(2)}_{n;(2,1)},\end{equation}
and therefore
\begin{equation} \bold{L}_{n;12}=\left[\begin{array}{cc}
    {M}^{(1)}_{n;(1,2)} & {M}^{(2)}_{n;(1,2)} \\
    -{M}^{(2)}_{n;(1,2)} & {M}^{(1)}_{n;(1,2)}
\end{array}\right]=\left[\begin{array}{cc}
    {M}^{(1)}_{n;(2,1)} & M^{(2)}_{n;(2,1)} \\
    -M^{(2)}_{n;(2,1)} & M^{(1)}_{n;(2,1)}
\end{array}\right]=\bold{L}_{n;21},\end{equation}
which is equivalent with $\bold{L}_{12}=\bold{L}_{21}$.

\section{Three terminal device with external magnetic field}
For a three terminal device with an external magnetic field, the two fluxes are a particle flux $\rho$ and a heat flux $q$. Each flux can be separated in two 'sub-fluxes': one from the first terminal to the second terminal and one from third terminal to the second terminal. The Onsager matrix is then given by
\begin{equation}\bold{L}=\frac{e^2T}{h}\int^{\infty}_{-\infty}dE\, F(E)\left[\begin{array}{cc}
    1 & \frac{E-\mu}{e}  \\
    \frac{E-\mu}{e} & \left(\frac{E-\mu}{e}\right)^2 
\end{array}\right]_{\rho,q}\otimes \left[\begin{array}{cc}
    1-|S_{11}(E,\bold{B})|^2 &-|S_{13}(E,\bold{B})|^2  \\
    -|S_{31}(E,\bold{B})|^2 & 1-|S_{33}(E,\bold{B})|^2
\end{array}\right]_{1,3},\end{equation}
with
\begin{equation}F(E)=\frac{1}{4k_BT cosh^2\left(\frac{E-\mu}{k_BT}\right)},\end{equation}
$k_B$ the Boltzmann constant, $\bold{S}(E,\bold{B})$ the scattering matrix of the first and third terminal, $B$ the magnetic field, $e$ the charge of one electron, and $\mu$ and $T$ are the chemical potential and the temperature of the second reservoir.
In particular, we see that
\begin{equation}\bold{L}_{\rho q}=\frac{e^2T}{h}\int^{\infty}_{-\infty}dE\, F(E)\frac{E-\mu}{e}\left[\begin{array}{cc}
    1-|S_{11}(E,\bold{B})|^2 &-|S_{13}(E,\bold{B})|^2  \\
    -|S_{31}(E,\bold{B})|^2 & 1-|S_{33}(E,\bold{B})|^2\end{array}\right]_{1,3}=\bold{L}_{q\rho},\end{equation}
which completes the proof.

One can also define a reduced Onsager matrix under the assumption that the net flux to the third terminal is zero, i.e.~$J_{\rho,32}=J_{q,32}=0$. This $2\times 2$ matrix  will no longer be symmetrical, as we will show now. To do this, we rearrange the fluxes $\bold{J}_1=(J_{\rho,12},J_{q,12})$ and $\bold{J}_3=(J_{\rho,32},J_{q,32})$, and the forces: $\bold{F}_1 =((T_1-T)/T^2,e(\mu_1-\mu)/T)$ and $\bold{F}_3=((T_3-T)/T^2,e(\mu_3-\mu)/T)$. The associated Onsager coefficients are given by
\begin{equation}\bold{L}_{ij}=\frac{e^2T}{h}\int^{\infty}_{-\infty}dE\, F(E)\left(\delta_{ij}-\left|S_{ij}(E,\bold{B})\right|^2\right)\left[\begin{array}{cc}
    1 & \frac{E-\mu}{e}  \\
    \frac{E-\mu}{e} & \left(\frac{E-\mu}{e}\right)^2 
\end{array}\right].\end{equation}
Note that the separate elements are symmetric, $\bold{L}_{ij}=\bold{L}_{ij}^T$, but that there is no direct connection between, $\bold{L}_{13}$ and $\bold{L}_{31}$. The constraint that the flux through the third terminal should be zero implies \begin{equation}\bold{J}_3=\bold{L}_{31}\bold{F}_1+\bold{L}_{33}\bold{F}_3=0,\end{equation}
or
\begin{equation}\bold{F}_3=-\bold{L}_{33}^{-1}\bold{L}_{31}\bold{F}_{1}.\end{equation}
Filing this in into the formula for the first flux gives
\begin{equation}\bold{J}_1=\left(\bold{L}_{11}-\bold{L}_{13}\bold{L}_{33}^{-1}\bold{L}_{31}\right)\bold{F}_1.\end{equation}
From here, it is clear that the reduced $2\times 2$ Onsager matrix, $\bold{L}'$, is given by
\begin{equation}\bold{L}'=\bold{L}_{11}-\bold{L}_{13}\bold{L}_{33}^{-1}\bold{L}_{31}.\end{equation}
Due to the symmetries of the Onsager matrices, we see that
\begin{equation}\bold{L}'^{T}=\bold{L}_{11}-\bold{L}_{31}\bold{L}_{33}^{-1}\bold{L}_{13}.\end{equation}
Combining these two equations, one concludes:
\begin{equation}\bold{L}-\bold{L}'^{T}=\bold{L}_{31}\bold{L}_{33}^{-1}\bold{L}_{13}-\bold{L}_{13}\bold{L}_{33}^{-1}\bold{L}_{31},\end{equation}
which is generally not equal to zero. Therefore, the reduced Onsager matrix is not symmetric, and the system no longer satisfies the second set of power-efficiency-dissipation relations.